\documentclass{iau}
\usepackage{graphicx}

\title[Magnetic fields in early-type stars] 
{Magnetic fields in early-type stars}

\author[Jason H. Grunhut \& Coralie Neiner]   
{Jason H. Grunhut$^1$
 \and Coralie Neiner$^2$}

\affiliation{$^1$European Southern Observatory, Karl-Schwarzschild-Str. 2, D-85748 Garching, Germany \\ email: {\tt jgrunhut@eso.org} \\[\affilskip]
$^2$LESIA, Observatoire de Paris, CNRS UMR 8109, UPMC, Universit\'e Paris Diderot, 5 place Jules Janssen, 92190, Meudon, France \\email: {\tt coralie.neiner@obspm.fr}}

\pubyear{2015}
\volume{305}  
\pagerange{119--126}
\jname{Polarimetry: From the sun to stars and stellar environments}
\editors{K.N. Nagendra, S. Bagnulo, R. Centeno, M. Mart{\'i}nez Gonz{\'a}lez}
\begin{document}

\maketitle

\begin{abstract}
For several decades we have been cognizant of the presence of magnetic fields in early-type stars, but our understanding of their magnetic properties has recently (over the last decade) expanded due to the new generation of high-resolution spectropolarimeters (ESPaDOnS at CFHT, Narval at TBL, HARPSpol at ESO). The most detailed surface magnetic field maps of intermediate-mass stars have been obtained through Doppler imaging techniques, allowing us to probe the small-scale structure of these stars. Thanks to the effort of large programmes (e.g. the MiMeS project), we have, for the first time, addressed key issues regarding our understanding of the magnetic properties of massive ($M > 8$\,$M_\odot$) stars, whose magnetic fields were only first detected about fifteen years ago. In this proceedings article we review the spectropolarimetric observations and statistics derived in recent years that have formed our general understanding of stellar magnetism in early-type stars. We also discuss how these observations have furthered our understanding of the interactions between the magnetic field and stellar wind, as well as the consequences and connections of this interaction with other observed phenomena.

\keywords{stars: magnetic fields, stars: early-type}
\end{abstract}

\firstsection 
\section{Magnetism in intermediate-mass AB stars}
In contrast to cool, late-type stars, early-type, higher-mass stars ($M\gtrsim1.5\,$M$_\odot$) have a significantly different internal structure. Instead of an outer convective envelope (as found in stars with $M\lesssim1.5\,$M$_\odot$), higher-mass stars host an outer radiative envelope. Without the necessary convective motion in their outer envelopes, higher-mass stars are therefore not expected to drive a dynamo. Indeed, classical observational tracers of dynamo activity fade and disappear among stars of spectral type F, at roughly the conditions predicting the disappearance of energetically-important envelope convection. Despite this fact, magnetic fields have been detected in a small population of (over 400) A- and B-type stars (e.g. \cite[Bychkov et al. 2009]{bychkov09}) dating back over 60 years to the first discovery by \cite[Babcock (1947)]{babcock47}. These Ap and Bp stars are easily identifiable as they present distinctive photospheric chemical peculiarities compared to non-magnetic A- and B-type stars, presumably a result of the magnetic field (e.g. \cite[Folsom et al. 2007]{folsom07}). While magnetism in low-mass stars is essentially ubiquitous, magnetic fields are only detected in 5-10\% of the population of intermediate-mass stars (e.g. \cite[Bagnulo et al. 2006]{bagnulo06}).

\section{Magnetism in massive OB stars}
The study of magnetism in massive stars ($M > 8$\,M$_\odot$) is still a relatively new field. This largely reflects the fact that, unlike the intermediate-mass magnetic stars that show strong and distinctive photospheric chemical peculiarities compared to non-magnetic A- and B-type stars (and therefore can be easily identified), the strong, radiatively-driven outflows of more massive stars and NLTE effects generally inhibit these chemical peculiarities or make them difficult to identify. Furthermore, with relatively few spectral lines from which to directly diagnose the presence of magnetism in the optical spectra of massive stars and additional complications such as emission or pulsations meant that these fields remained undetected by previous generations of instrumentation.

The large majority of massive stars with known magnetic fields belong to the He-peculiar early B-type stars (e.g. \cite[Bohlender et al. 1987]{bohlender87}), such as the archetype magnetic B2 star $\sigma$~Ori~E (\cite[Landstreet \& Borra 1978]{landstreet78}). These stars represent the high-mass extension of the chemically-peculiar intermediate-mass stars. The first genuinely massive ($>10\,M_\odot$) magnetic stars were only detected about 15 years ago (e.g. the classical B1III pulsator $\beta$ Cep (\cite[Henrichs et al. 2000, 2013]{henrichs00,henrichs13}; \cite[Donati et al. 2001]{donati01}); the young massive O7V star $\theta^1$~Ori~C (\cite[Donati et al. 2002]{donati02})). Prior to 2009, only 3 magnetic O-type stars were even known ($\theta^1$~Ori C; the Of?p star HD\,191612 (\cite[Donati et al. 2006]{donati06a}); the evolved O9.7Ib binary star $\zeta$~Ori~A (\cite[Bouret et al. 2008]{bouret08})), and 9 of the 13 of the most massive stars were discovered between 2005 and 2009. These breakthroughs were a consequence of a new generation of high-efficiency, broadband spectropolarimeters mounted on medium-class telescopes and new techniques used to increase the signal-to-noise ratio, such as Least-Squares Deconvolution (LSD; \cite[Donati et al. 1997]{donati97}).

In 2008, the Magnetism in Massive Stars (MiMeS) international project was set up to investigate the magnetic properties and related physics of massive stars. Observations were gathered over 5 years (2008-2012) with the three high-resolution spectropolarimeters available in the world: ESPaDOnS at CFHT in Hawaii (PI: G. Wade), Narval at TBL in France (PI: C. Neiner), and HarpsPol at ESO in La Silla in Chile (PI: E. Alecian). 

The MiMeS project included a spectropolarimetric survey of $\sim$550 O and early B stars, aimed at determining the statistical incidence fraction of magnetic fields in this class of stars and at investigating the connection between magnetism and various observed phenomena. 7$\pm$1\% of the massive stars observed within the MiMeS survey were found to be magnetic, with a detection threshold of a few tens of gauss. This is fully consistent with the fraction of magnetic stars detected among intermediate-mass stars (\cite[e.g. Power et al. 2007]{power07}). However, no correlation was found between the presence of a magnetic field, or its strength, and fundamental stellar parameters, such as temperature, mass, or rotation.

The MiMeS survey has had a significant impact on the field of massive star magnetism. Of particular interest, the survey has more than tripled the number of known massive O-type stars with detectable magnetic fields. From these stars, the MiMeS project was able to establish the connection between the presence of magnetic fields and the Of?p classification (\cite[Walborn 1972]{walborn72}; \cite[Walborn et al. 2010]{walborn10}). All of the known Galactic stars of the Of?p class were observed and fields were either detected or confirmed (see e.g. \cite[Wade et al. 2014b]{wade14b}, and references therein). As there are now 5 known extra-Galactic Of?p stars (\cite[Walborn et al. 2010]{walborn10}; \cite[Massey et al. 2014]{massey14}), this therefore implies the identification of the first magnetic stars outside of the Galaxy.

\section{Field characteristics}
\begin{figure}[h]
\begin{center}
\includegraphics[width=5.2in]{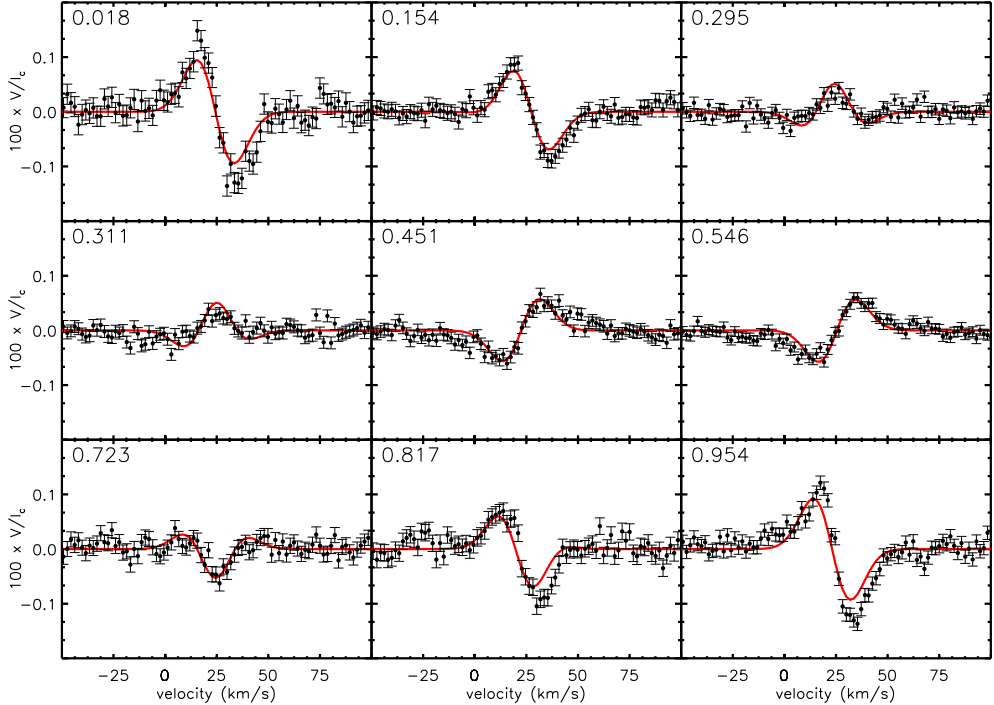} 
\caption{Example LSD Stokes $V$ profiles (circles) for HD\,57682 from \cite[Grunhut et al. (2012b)]{grunhut12c}, with error bars representing the 1$\sigma$ uncertainties of each pixel. Rotational phases are indicated in the upper left of each panel. Also shown is the best-fit model to these profiles (solid line; surface dipole field strength of 700\,G and magnetic obliquity of 68$^\circ$, for an assumed inclination of 60$^\circ$).}
\label{fig1}
\end{center}
\end{figure}

The overwhelming majority of higher-mass stars present globally-ordered, topologically simple magnetic fields, often characterized by a dipole or a low-order multipole (e.g. \cite[Auri{\` e}re et al. 2007]{auriere07}), with the magnetic axis oblique to the rotation axis. The majority of these studies have characterized the magnetic topology using only the net, surface integrated, line-of-sight component of the magnetic field (the longitudinal field $B_\ell$) that is probed via circular polarization. Zeeman Doppler Imaging (ZDI) based on intensity and circular polarization (Stokes $V$) profiles have been performed for several B and A stars and allow one to reveal the detailed structure of the magnetic field on the stellar surface. These magnetic maps support the $B_\ell$ measurements and that massive and intermediate-mass stars are found to be mostly dipolar, although a few examples of more complex fields exist (e.g. HD\,37776; \cite[Kochukhov et al. 2011]{kochukhov11}). Adding information from linear polarization (Stokes $QU$) to ZDI mapping shows additional small-scale structures and deviations from a pure dipole (e.g. \cite[Kochukhov \& Wade 2010]{kochukhov10}; see also Ros\'en et al., these proceedings). The ZDI technique has not been applied to O stars yet, but parametrized models fitted to Stokes profiles confirm that the field of O stars are also mostly dipolar (see Fig.~\ref{fig1}). 

The fields observed in O, B and A stars are rather strong. For both intermediate-mass and massive stars, their polar field strengths are typically of the order of 1000 G and range from 100 G to a few tens of kG, as shown by \cite[Auri{\` e}re et al. (2007)]{auriere07} for intermediate-mass stars and the MiMeS collaboration (\cite[Wade et al. 2014a]{wade14a}) for massive stars. In addition, the 100 G limit under which basically no magnetic field is found is well above the detection threshold of a few tens of gauss reached in these studies. A magnetic desert thus seems to exist below this field limit. Note, however, that a new class of ultra-weak fields has recently been discovered below 1 G (e.g. \cite[Petit et al. 2011; Ligni\`eres et al. 2014; see also Blazere et al., these proceedings]{petit11,lignieres14}).

Contrary to cooler stars, the magnetic fields of hot stars are stable. The first magnetic intermediate-mass stars were discovered $\sim$60 years ago and their configuration show no variation over this timeframe, except for rotational modulation due to the obliquity of the magnetic field versus the rotation axis. Surface field maps of hot stars obtained decades apart are very similar (\cite[e.g. Silvester et al. 2014]{silvester14}). The first examples of magnetic massive stars were discovered in the late 90's, and therefore the time span of observations is shorter. However, even from this limited baseline, these targets also show remarkably stable fields. This is best confirmed by proxies related to the magnetic field, such as H$\alpha$ or X-ray emission produced in the magnetosphere or the confined wind lines observable in the UV (\cite[see e.g. Grunhut et al. 2012b]{grunhut12c} and further discussed below).

\section{Magnetically confined winds and magnetospheres}
The study of magnetically confined winds still remains active even after almost four decades since the initial discovery of a magnetic field in the early B-type star $\sigma$~Ori~E (\cite[Landstreet \& Borra 1978]{landstreet78}), and the realization that the interaction of this field with the radiatively-driven stellar wind was responsible for a multitude of observational phenomena in this and other magnetic early-B stars (e.g. \cite[Pedersen \& Thomsen 1977]{pedersen77}; \cite[Walborn 1982]{walborn82}; \cite[Groote \& Hunger 1982]{groote82}; \cite[Nakajima 1985]{nakajima85}; \cite[Shore \& Brown 1990]{shore90}; \cite[Leone \& Umana 1993]{leone93}; \cite[Babel \& Montmerle 1997a]{babel97a}). In fact, building on the earlier work in this field (e.g. \cite[Nakajima 1985]{nakajima85}; \cite[Shore 1987]{shore87}; \cite[Babel \& Montmerle 1997a]{babel97a}), a new generation of numerical and semi-analytical magnetohydrodynamical (MHD) models (e.g. \cite[ud-Doula \& Owocki 2002]{uddoula02}; \cite[Townsend \& Owocki 2005]{townsend05}; \cite[Sundqvist et al. 2012]{sundqvist12}) have led the advancement on the theoretical front over the last 10-15 years. These models have not only proven to show excellent agreement with observations, but have also inspired new generations of observational studies to investigate their predictions.

The common picture is that radiation drives wind plasma from the stellar surface from the footpoints of magnetic field lines to the loop tops. As the plasma collides and cools, it generates detectable and variable X-ray emission from this shock region (e.g. \cite[Babel \& Montmerle 1997b]{babel97b}; \cite[Gagn{\'e} et al. 2005]{gagne05}). As the post-shock material continues to cool, it can form a magnetosphere that generates emission and is often observable in optical hydrogen recombination lines  (e.g. \cite[Howarth et al. 2007]{howarth07}; \cite[Bohlender \& Monin 2011]{bohlender11}).

The wind should be confined if the local magnetic energy density is stronger than the local wind energy density, as characterized by the magnetic confinement parameter (\cite[ud-Doula \& Owocki 2002]{uddoula02}) $\eta_* = B_{\rm eq}^2 R_\star^2/\dot{M}v_\infty$, which depends on the star's equatorial surface field ($B_{\rm eq}$; for a dipole field, this is half the strength of the field at the pole), the stellar equatorial radius $R_\star$, and the wind terminal momentum ($\dot{M}v_\infty$, for the wind feeding rate $\dot{M}$, i.e. the mass-loss the star would have if no magnetic field was present, and the wind terminal velocity $v_\infty$). Plasma is expected to be confined out to a distance where the wind energy density is balanced by the magnetic energy density, called the Alfv{\' e}n radius, given by $R_A/R_\star \sim 0.3 + (\eta_* + 0.25)^{1/4}$ for a dipole field (\cite[ud-Doula \& Owocki 2002]{uddoula02}). Beyond this distance, the wind dominates and the closed magnetic field lines are dragged with the wind and stretched to open field lines.

If the star is undergoing rapid rotation, then the wind plasma is centrifugally supported and magnetically confined into dense regions (or clouds) along gravito-centrifugal potential minima, which are forced to co-rotate with the host star (e.g. \cite[Shore \& Brown 1990]{shore90}; \cite[Townsend \& Owocki 2005]{townsend05}; \cite[ud-Doula, Owocki \& Townsend 2008]{uddoula08}). Centrifugal confinement of the plasma occurs beyond the Kepler, or co-rotation, radius $R_{\rm k}= 3/2\omega^{-2/3}R_{\rm p}$ (\cite[ud-Doula, Owocki \& Townsend 2008]{uddoula08}), where $\omega$ is the rotational frequency of the star and $R_{\rm p}$ is the polar radius. Below this distance, magnetically confined plasma is expected to fall back onto the star on free-fall timescales (\cite[ud-Doula \& Owocki 2002]{uddoula02}), while beyond this distance the plasma is centrifugally supported. If $R_{\rm A}>R_{\rm K}$ then the gravito-centrifugally supported plasma accumulates, forming magnetospheric clouds. 

Recently, \cite[Petit et al. (2013)]{petit13} constructed a rotation-confinement diagram that establishes essentially two types of magnetospheres: the centrifugally supported magnetospheres (CMs) belonging to stars with rapid rotation that satisfy $R_{\rm A} > R_{\rm K}$; or the dynamical magnetospheres that satisfy $R_{\rm K} > R_{\rm A}$. Observations of stars with CMs or DMs have established some basic characteristic of the hydrogen Balmer emission: broad emission features are found in stars with CMs that are most often found at high velocities, well outside the projected rotational velocity ($v \sin i$) of the star. On the other hand, DMs show much narrower emission that is confined within the $v \sin i$ of the star. Emission is typically observed from DMs only around  O-type stars, as their wind feeding rates are sufficiently high to produce magnetospheres with a statistically overdense region centered around the magnetic equator. Observations and theory show that the magnetospheres are seen in maximum emission when projected on the sky to their fullest extent, and in minimum emission (or even possibly absorption for CMs, if they occult the star) when viewed edge-on. 

Detailed observations of CM hosting stars have also revealed the effects of the magnetospheres on the line-profile-variations (LPV; e.g. \cite[Oksala et al. 2012]{oksala12}; \cite[Grunhut et al. 2012a]{grunhut12b}; \cite[Rivinius et al. 2013]{rivinius13a}).  As illustrated by several recent studies (e.g. \cite[Townsend, Owocki \& Groote 2005]{townsend05b}; \cite[Bohlender \& Monin 2011]{bohlender11}; \cite[Grunhut et al. 2012a]{grunhut12b}; \cite[Rivinius et al. 2013]{rivinius13a}), the LPV show specific characteristics that reflect the rigid motion of these clouds that is unique to these stars. The characteristics of these clouds (e.g. emission strength, location, extent) are in generally good agreement with models (\cite[Townsend, Owocki \& Groote 2005]{townsend05}).

\begin{figure}[h]
\begin{center}
\includegraphics[width=2.6in]{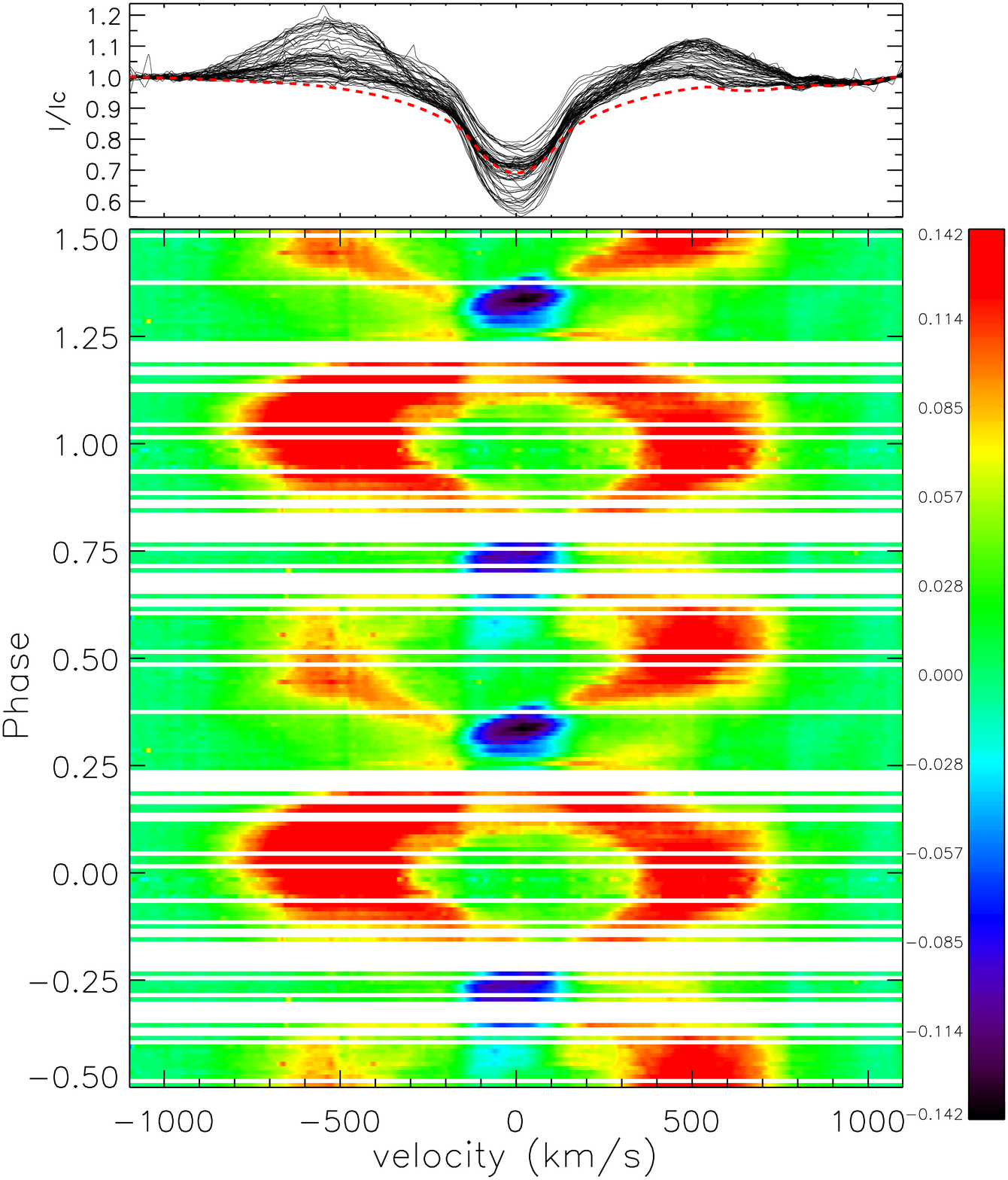} 
\includegraphics[width=2.6in]{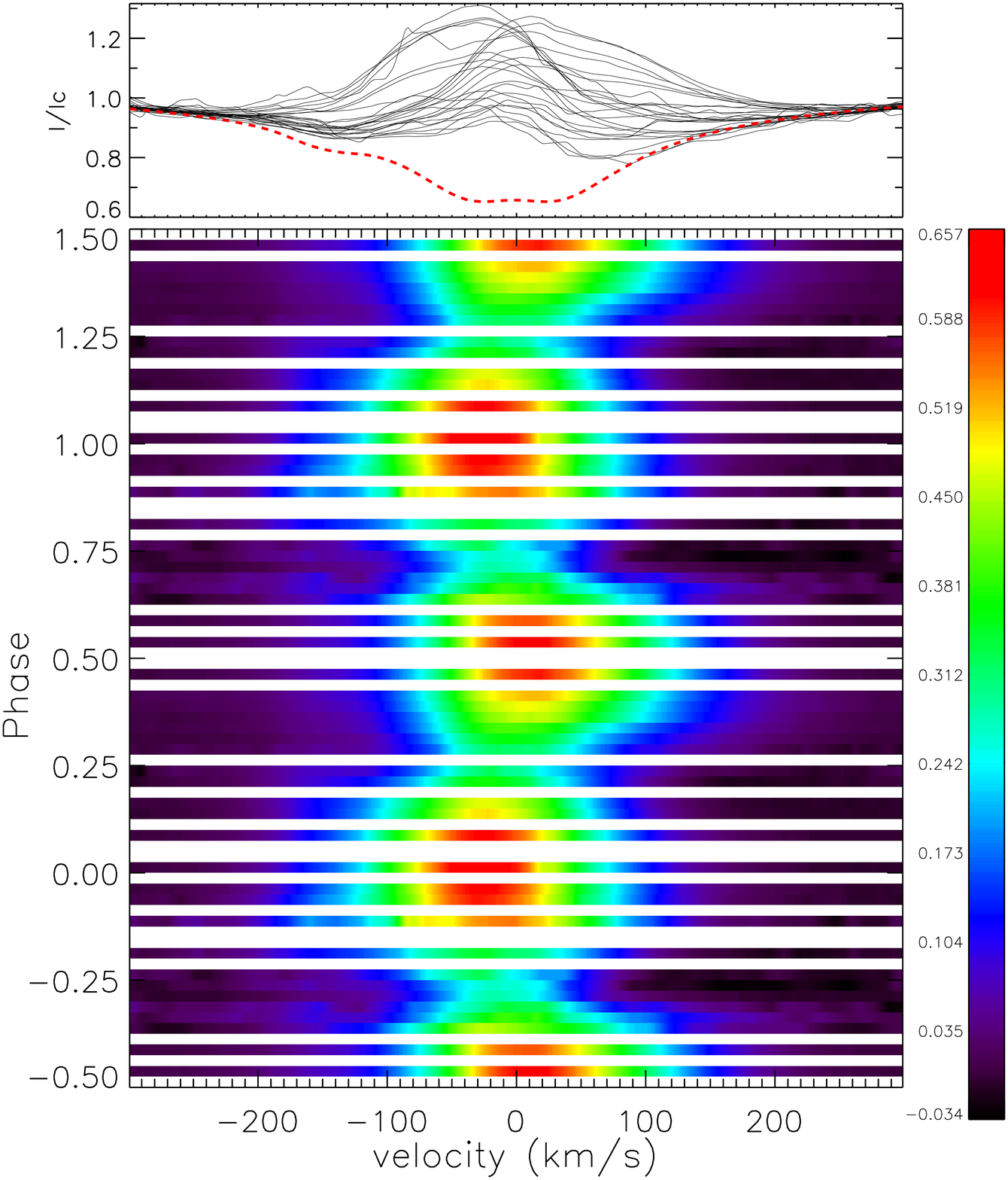}
\caption{Comparison of dynamic spectra showcasing the differences in the emission variability found in the H$\alpha$ line for the CM hosting star $\sigma$ Ori E (left) and the DM hosting star HD\,57682 (right), with similar magnetic geometries. The differences are shown relative to theoretical spectra to highlight the relative contribution of the magnetosphere to the emission properties. While both show rotational modulation, the emission variations in the CM traces the motions of cloud-like features that extend out to high $v\sin i$, while the DM only shows variations in the intensity of the line profile.}
\label{dyn_fig}
\end{center}
\end{figure}

While the emission features corresponding to CMs characterize their position and motion, the same does not occur for DMs. The emission features observed in these stars still undergo rotational modulation, but the pattern of variability shows a systematic change in intensity. The models of \cite[Sundqvist et al. (2012)]{sundqvist12} and \cite[Grunhut et al. (2012b)]{grunhut12c} (using the results of two-dimensional MHD simulations) are in excellent agreement with observations, although these models required the introduction of extra turbulence to match the width of the emission lines. A side-by-side comparison of the LPV resulting from a star with a CM versus a DM is presented in Fig.~\ref{dyn_fig}.

\section{The connection between magnetic fields and other phenomena}
\subsection{Be stars and decretion disks}

Magnetic fields have been proposed as an explanation for the decretion disk surrounding classical Be stars (see \cite[Cassinelli \& Neiner 2005]{cassinelli05}). For example, the magnetically torqued disk model (\cite[Cassinelli, Brown \& Maheswara 2002]{cassinelli02}) is one proposed scenario. 

Claims that a significant fraction of classical Be stars may indeed be magnetic have been put forward based on low-resolution spectropolarimetric studies using FORS (\cite[Hubrig, North \&  Sch{\"o}ller 2007]{hubrig07}; \cite[Hubrig et al. 2009]{hubrig09b}; \cite[Yudin et al. 2011]{yudin11}). Although these results were later shown to be spurious (\cite[Bagnulo et al. 2012]{bagnulo12}), they nonetheless motivated further study. The MiMeS survey included $\sim$85 classical Be stars selected from the BeSS catalog (\cite[Neiner et al. 2011]{neiner11}) according to the criteria of \cite[Porter \& Rivinius (2003)]{porter03} - as early-type emission line stars with Keplerian disks - to distinguish these stars from other emission line stars whose emission is of a different origin (e.g. accretion disks from Herbig stars). The observations resulted in no direct evidence of a magnetic field in any of these stars (\cite[Wade et al. 2014]{wade14b}), despite the fact that the survey achieved a magnetic sensitivity similar to that of the larger sample.

Periodic modulation of UV and visible lines in the classical Be star $\omega$\,Ori (\cite[Neiner et al. 2003; Neiner et al. 2012]{neiner03c,neiner12d}) has been interpreted as indirect evidence for a magnetic field. The study of \cite[Neiner et al. (2012)]{neiner12d} could not rule out the possibility of a weak magnetic field in this star and therefore concluded that weak magnetic fields might still be present in some classical Be stars, but, as argued by \cite[ud-Doula \& Owocki (2002)]{uddoula02}, they are not at the origin of the circumstellar decretion disk. Classical Be star disks are in Keplerian orbits in the rotational equator of the star, contrary to magnetospheres which corotate closer to the star (\cite[Neiner et al. 2003]{neiner03c}). 

\subsection{Nitrogen enrichment}
Rotationally-induced mixing that brings fusion products from the core to the surface during the hydrogen burning phase is well supported by theory (e.g. \cite[Brott et al. 2011]{brott11}) and observations (\cite[e.g. Hunter et al. 2008]{hunter08}). The result of this mixing is usually nitrogen-enhancement at the surface of these stars. However, a surprising result from observational studies is the discovery of a small population of stars with low projected rotational velocities, yet nitrogen enhancement (\cite[Morel et al. 2006]{morel06}; \cite[Hunter et al. 2008]{hunter08}). The origin of this enhancement is not understood but may be related to binarity (\cite[Langer 2008]{langer08}), pulsations (\cite[Aerts et al. 2014]{aerts14}), or enhanced mixing due to magnetic breaking (\cite[Meynet, Eggenberger \& Maeder 2011]{meynet11}).

Nineteen stars with well determined nitrogen abundances (\cite[Nieva \& Przybilla 2012]{nieva12}) have been observed within the MiMeS project and a follow-up study of sharp-lined B-type stars to investigate the connection between magnetism and nitrogen enrichment (\cite[Wade et al. 2015]{wade15a}). The results from this study indicate that all of the magnetic stars (3) show significant enhancement, although several of the other nitrogen enriched stars are not found to be magnetic to a high precision. While the size of this sample is relatively small, there is no reason to favour that magnetic breaking is responsible for enhancing the surface abundance of nitrogen in magnetic stars. It is more likely that these stars are enriched due to the same mechanism that affects the other stars that show no evidence for strong, globally-ordered magnetic fields. 

\section{Summary}
It is now well-established that large-scale magnetic fields are detected in a small fraction (5-10\%) of early-type OBA stars. The similar statistics and magnetic field characteristics of these stars during the pre-MS (\cite[Alecian et al. 2013]{alecian13a}) and MS phases suggest that the physics of magnetism in stellar radiative zones remains unchanged over millions of years of evolution and across more than an order of magnitude of mass. The substantial contrast relative to late-type, low-mass stars suggests a fundamentally different origin for the fields in early-type stars (see Neiner et al., these proceedings for further discussion). 




\end{document}